\documentclass[journal,twocolumn,10pt]{IEEEtran}
\usepackage{graphicx}
\usepackage{amssymb}
\usepackage{amsmath}
\usepackage{mathtools}
\usepackage{dsfont}
\usepackage{cite}
\usepackage{stfloats}
\usepackage{subfigure}
\usepackage{psfrag}
\usepackage[mathscr]{euscript}
\usepackage{acronym}  
\usepackage{booktabs}
\usepackage{bbm}

\usepackage{float}
\usepackage{balance}

\acrodef{CCDF}{complementary cumulative distribution function}
\acrodef{CF}{characteristic function}
\acrodef{PPP}{Poisson point processe}
\acrodef{RV}{random variable}
\acrodef{i.i.d.}{independent and identically distributed}
\acrodef{PDF}{probability distribution function}
\acrodef{CDF}{cumulative distribution function}
\acrodef{ch.f.}{characteristic function}
\acrodef{AWGN}{additive white Gaussian noise}
\acrodef{SNR}{signal-to-noise ratio}
\acrodef{LRT}{likelihood ratio test}
\acrodef{DRT}{distance ratio test}
\acrodef{GLRT}{generalized likelihood ratio test}
\acrodef{CRLB}{Cram\'{e}r-Rao lower bound}
\acrodef{CRB}{Cram\'{e}r-Rao bound}
\acrodef{ZZLB}{Ziv-Zakai lower bound}
\acrodef{ZZB}{Ziv-Zakai bound}
\acrodef{LOS}{line-of-sight}
\acrodef{ToF}{time-of-flight}
\acrodef{NLOS}{non-line-of-sight}
\acrodef{GDOP}{geometric dilution of precision}
\acrodef{GPS}{Global Positioning System}
\acrodef{FIM}{Fisher information matrix}
\acrodef{PEB}{position error bound}
\acrodef{SPEB}{squared position error bound}
\acrodef{TOA}{time-of-arrival}
\acrodef{TOF}{time-of-flight}
\acrodef{WSN}{wireless sensor network}
\acrodef{MAC}{medium access control}
\acrodef{RSS}{received signal strength}
\acrodef{WAF}{wall attenuation factor}
\acrodef{TDOA}{time difference-of-arrival}
\acrodef{RF}{radiofrequency}
\acrodef{RTT}{round-trip time}
\acrodef{AOA}{angle-of-arrival}
\acrodef{MF}{matched filter}
\acrodef{ED}{energy detector}
\acrodef{ML}{maximum likelihood}
\acrodef{MSE}{mean-square error}
\acrodef{RMSE}{root-mean-square error}
\acrodef{LEO}{localization error outage}
\acrodef{ppm}{part-per-million}
\acrodef{ACK}{acknowledge}
\acrodef{UWB}{Ultrawide bandwidth}
\acrodef{TNR}{threshold-to-noise ratio}
\acrodef{LS}{least squares}
\acrodef{IR-UWB}{impulse radio UWB}
\acrodef{FCC}{Federal Communications Commission}
\acrodef{TH}{time-hopping}
\acrodef{PPM}{pulse position modulation}
\acrodef{MUI}{multi-user interference}
\acrodef{PDP}{power delay profile}
\acrodef{BPZF}{band-pass zonal filter}
\acrodef{SIR}{signal-to-interference ratio}
\acrodef{SINR}{signal-to-interference-plus-noise ratio}
\acrodef{RFID}{radio frequency identification}
\acrodef{WPAN}{wireless personal area network}
\acrodef{WWB}{Weiss-Weinstein bound}
\acrodef{DP}{direct path}
\acrodef{MF}{matched filter}
\acrodef{MMSE}{minimum-mean-square-error}
\acrodef{SBS}{serial backward search}
\acrodef{SBSMC}{serial backward search for multiple clusters}
\acrodef{NBI}{narrowband interference}
\acrodef{WBI}{wideband interference}
\acrodef{INR}{interference-to-noise ratio}
\acrodef{CR}{channel response}
\acrodef{CIR}{channel impulse response}
\acrodef{CR}{channel  response}
\acrodef{RADAR}{radar}
\acrodef{MUR}{Multistatic radar}
\acrodef{JBSF}{jump back and search forward}
\acrodef{HDSA}{high-definition situation-aware}
\acrodef{RRC}{root raised cosine}
\acrodef{ST}{simple thresholding}
\acrodef{BTB}{Bellini-Tartara bound}
\acrodef{P-Max}{$P$-Max}  
\acrodef{MIMO}{multiple-input multiple-output}
\acrodef{MAP}{maximum a posteriori}
\acrodef{FG}{factor graph}
\acrodef{OP}{outage probability}
\acrodef{WED}{wall extra delay}
\acrodef{RMS}{root mean square}
\acrodef{SPAWN}{sum-product algorithm over a wireless network}
\acrodef{MDD}{minimum distance distribution}
\acrodef{MAP}{maximum a posteriori probability}
\acrodef{SAP}{small cell access point}
\acrodef{UE}{user equipment}
\acrodef{MBS}{macro cell base station}
\acrodef{UER}{\ac{UE} Relay}
\acrodef{D2D}{device-to-device}
\acrodef{MBS}{macro base station}
\acrodef{CSI}{channel state information}
\acrodef{OGR}{outage guard region}
\acrodef{FUR}{feasible UER region}
\acrodef{EHR}{energy harvesting region}
\acrodef{EH}{energy harvesting}
\acrodef{D2D-EHSN}{D2D communication provided \ac{EH} small cell network}
\acrodef{D2D-EHHN}{D2D communication provided \ac{EH} heterogeneous network}
\acrodef{3GPP}{3rd Generation Partnership Project}
\acrodef{BS}{base station}
\acrodef{DF}{decode and forward}
\acrodef{CCDF}{complementary cumulative distribution function}
\acrodef{ZF}{zero forcing}
\acrodef{RZF}{regularized zero forcing}
\acrodef{WLLN}{weak law of large number}
\acrodef{SLLN}{strong law of large numbers}
\acrodef{TDD}{Time-division duplex}
\acrodef{EE}{energy efficiency} 
\acrodef{HetNet}{heterogeneous network} 
\acrodef{SCP}{Single Cell Processing}
\acrodef{CBF}{Coordinated Beamforming}
\usepackage{color}
\usepackage{dsfont}
\usepackage{bbm}

\def\PMT{P_{\mathrm{mt}}}

\def\LB{\lambda_{\mathrm{b}}}

\def\LU{\lambda_{\mathrm{u}}}

\DeclareMathAlphabet{\mathsf}{OML}{cmbr}{m}{it}

\newtheorem{definition}{\bf Definition}
\newtheorem{theorem}{\bf Theorem}
\newtheorem{lemma}{\bf Lemma}
\newtheorem{corollary}{\bf Corollary}

\newtheorem{remark}{\bf Remark}

\newtheorem{assumption}{\bf Assumption}

\newcommand{\bd}{\begin{description}}
\newcommand{\ed}{\end{description}}
\newcommand{\be}{\begin{enumerate}}
\newcommand{\ee}{\end{enumerate}}
\newcommand{\bi}{\begin{itemize}}
\newcommand{\ei}{\end{itemize}}
\newcommand{\bl}{\begin{list}}
\newcommand{\el}{\end{list}}
\newcommand{\bt}{\begin{tabbing}}
\newcommand{\et}{\end{tabbing}}

\newcommand{\paperTitle}{SIR Coverage Analysis in Cellular Networks with Temporal Traffic: A Stochastic Geometry Approach}

\begin{document}

{
\title{\paperTitle}

\author{

	    Howard~H.~Yang, \textit{Member, IEEE},
        and Tony~Q.~S.~Quek, \textit{Fellow, IEEE}


\thanks{H.~H.~Yang and T.~Q.~S.~Quek are with the Information Systems Technology and Design Pillar, Singapore University of Technology and Design, Singapore (e-mail: howard\_yang@sutd.edu.sg, tonyquek@sutd.edu.sg).}
}
\maketitle
\acresetall
\thispagestyle{empty}
\begin{abstract}
The bloom in mobile applications not just bring in enjoyment to daily life, but also imposes more complicated traffic situation on wireless network. A complete understanding of the impact from traffic profile is thus essential for network operators to respond adequately to the surge in data traffic.
In this paper, based on stochastic geometry and queuing theory, we develop a mathematical framework that captures the interplay between the spatial location of base stations (BSs), which determines the magnitude of mutual interference, and their temporal traffic dynamic.
We derive a tractable expression for the SIR distribution, and verify its accuracy via simulations.
Based on our analysis, we find that
$i$) under the same configuration, when traffic condition changes from light to heavy, the corresponding SIR requirement can differ by more than 10~dB for the network to maintain coverage,
$ii$) the SIR coverage probability varies largely with traffic fluctuation in the sub-medium load regime, whereas in scenario with very light traffic load, the SIR outage probability increases linearly with the packet arrival rate,
$iii$) the mean delay, as well as coverage probability of cell edge user equipments (UEs) are vulnerable to the traffic fluctuation, thus confirms its appeal for traffic-aware communication technology.
\end{abstract}

\acresetall

\section{Introduction}\label{sec:intro}
The rapid evolution of mobile applications imposes more complicated traffic condition on wireless networks, where not only the data demand grows exponentially \cite{LopDinCla:15}, but more importantly, the the content is largely changing from mobile voice to multimedia \cite{niu2011tango}. To give an adequate response to the surge in mobile data traffic, network operators need a complete understanding on the impact of temporal traffic.
In this article, we aim to evaluate how the traffic statistic affects the wireless networks, and to find those aspects that are most vulnerable.

\subsection{Background and Related Work}
Due to the broadcast nature of wireless channel, transmitters in space sharing a common spectrum will interact with each other through the interference they cause. To characterize the performance of such networks, stochastic geometry has been recently introduced as a way to assess performance of wireless links in large-scale networks \cite{HaeAndBac:09,Hae:12,BacBla:09,ElSHosHae:13,ElSSulAlo:17,DiLuGua:16}.
The intrinsic elegance in modeling and analysis has popularized its application in evaluating performance among various wireless systems, including ad-hoc networks \cite{HaeAndBac:09}, cellular networks \cite{AndBacGan:11}, or more advanced heterogeneous networks \cite{DhiGanBac:12}, even with device-to-device (D2D) communication \cite{YanLeeQue:16} and multiple-input multiple-output (MIMO) technology \cite{YanGerQue:16,YanGerQue:17}. However, the main drawback of these models is that they heavily rely on the \textit{full buffer} assumption, i.e., every link always has a packet to transmit, and do not allow one to represent random traffic. While the additional dimension of randomness in temporal domain increases the complexity in analysis, it is nevertheless a crucial factor in understanding network performance, especially for the next generation wireless system that faces more voliated traffic conditions \cite{ZhoHaeZhe:16,AndBuzCho:14}.

The main difficulty with queuing in wireless network comes from the interdependency among the evolution of different queues, which is usually referred to as interacting queues \cite{rao1988stability}. Because of interference, the queue status of one transmitter can affect, and also be affected by, the queue status of its neighbors, hence making the analysis very difficult.
Conventionally, the queuing interaction through wireless medium is studied using simple collision models \cite{BerGalHum:92,rao1988stability,LuoEph:99,BorMcDPro}.
In such models, discrete time ALOHA protocol is usually employed, where each of the $N$ terminals initiates a transmission attempt at every slot: If more than two terminals transmit simultaneously, a collision occurs and all the terminals retransmit their packets in next slot with the same probability \cite{BerGalHum:92}.
Analytical results about system stability can be obtained via exact form in scenarios with few (two or three) transmitters \cite{rao1988stability}, or through approximations in asymptotic regime with infinitely many transmitters \cite{BorMcDPro:12}.
However, these models over simplify the wireless channel and lack the ability to tract the interference, which differs according to distance as well as channel gains, thus do not capture the information-theoretic interactions precisely. Recent attempts to address this issue are made in \cite{ZhoQueGe:16,GhaElsBad:17,yang2017packet,ZhoHaeQue:16,GeoSpyKal:17}, where queuing theory is combined with stochastic geometry to model the dynamic from both temporal and spatial domains. The results provide the necessary and sufficient conditions for network to be stable \cite{ZhoHaeQue:16}, and different performance metrics, including  success transmission probability \cite{GhaElsBad:17}, delay \cite{ZhoQueGe:16}, and packet throughput \cite{yang2017packet} have been subsequently derived.
While giving more refined analysis, these results either provide only bounds that are not necessary tight \cite{ZhoQueGe:16,ZhoHaeQue:16}, or are only appliable to networks with light traffic \cite{yang2017packet}.
The most related work is from \cite{GhaElsBad:17}, where the authors applied Geo/PH/1 queuing model to account for the interference-based queues among UEs and analyzed the SIR performance under three different transmission schemes.
However, the requirement for full channel inversion limits its generalization, and the restricted stable region constrains its application to relatively low traffic condition with small SIR detection threshold, and thus prevents one to take a complete treatment on traffic statistic.
To this end, a mathematical framework that captures the spatial--temporal dynamic of the network, and adapts to scenarios with different traffic conditions is of necessity to be explored.

\subsection{Approach and Summary of Contributions}
In this paper, we model the BS deployment and UE locations as independent Poisson point processes (PPPs), where each BS maintains an infinite capacity buffer to store the incoming packets.
The queuing dynamic is modelled via a descrete time system, where we consider the arrival of packets at each BS to be independent Bernoulli process. By combining stochastic geometry with queuing theory, we obtain a tractable expression for the SIR coverage probability.
With the developed framework, we can explicitly characterize the SIR variation due to change of traffic condition, and its consequential impact on system stability, as well as delay distribution. Our main contributions are summarized below.

\begin{itemize}
\item We develope a mathematical framework that captures the interplay between the spatial geometry of wireless links and their temporal traffic dynamic. Our analysis is tractable, and takes into account all the key features of a cellular network, including traffic profile, small-scale fading and path loss, random network topology, and queuing interaction.
\item Unlike \cite{GhaElsBad:17}, our result not only provides the standard SIR coverage probability, but also gives a more precise description about the fraction of UEs achieving SIR at different levels. For instance, the SIR coverage probability of cell-edge UEs can be easily derived via our result.
\item We discuss the sufficient and necessary conditions for the network to be stable, and provide an approximation for the stable region. We also derive the mean delay distribution, by accounting for both queuing and transmission delay.
\item Using the developed analysis, we find that under the same network configuration, there is more than 10~dB SIR difference between light and heavy traffic conditions.
Moreover, in the very light traffic regime, the network SIR outage probability is shown to increase linearly with packet arrival rate. The result also reveals that the mean delay, as well as cell-edge UE rate, are vulnerable to the variation of traffic condition, hence urging advanced solution to adapt with traffic profile.
\end{itemize}

The remainder of the paper is organized as follows.
We introduce the system model in Section~\ref{sec:sysmod}.
In Section~\ref{sec:Analysis}, we detail the analysis of SIR distribution in cellular networks with temporal traffic.
We show the simulation and numerical results in Section~\ref{sec:NumAnal}, that confirm the accuracy of our analysis, and provide insights about the impact of traffic profile on network performance. We conclude the paper in Section~\ref{sec:conclusions}.

\section{System Model}\label{sec:sysmod}
In this section, we provide a general introduction to the network topology, the traffic profile, as well as the propagation and failure retransmission model.
The main notations used throughout the paper are summarized in Table~I.

\begin{table}
\caption{Notation Summary
} \label{table:notation}
\begin{center}
\renewcommand{\arraystretch}{1.3}
\begin{tabular}{c  p{5.5cm} }
\hline
 {\bf Notation} & {\hspace{2.5cm}}{\bf Definition}
\\
\hline
$\Phi_{\mathrm{b}}$; $\LB$ & PPP modeling the location of \acp{BS}; \ac{BS} deployment density \\
$\Phi_{\mathrm{u}}$; $\LU$ & PPP modeling the location of UEs; UE deployment density \\
$\PMT$; $\alpha$ & BS transmit power; path loss exponent \\
$\mathcal{S}_{\varepsilon}$ & $\varepsilon$-stable region of wireless network, under which the fraction of unstable queues is less than $\varepsilon$ \\
$\xi$; $\xi_{\mathrm{c}, \varepsilon}$ & Packet arrival rate; critical arrival rate for $\varepsilon$-stability \\
$\gamma_{x_0, t}$; $\theta$ &  Received SIR of typical UE at time slot $t$; SIR decoding threshold \\
$\mu_{x_0, t}^\Phi$ & Conditional SIR coverage probability at time slot $t$ \\
\hline
\end{tabular}
\end{center}\vspace{-0.63cm}
\end{table}%

\subsection{Network Topology and Traffic Model}
We consider the downlink of a cellular network, as depicted in Fig.~\ref{fig:TrafficModel}, that consists of randomly deployed \acp{BS} whose spatial locations follow independent Poisson point process (PPP) $\Phi_{\mathrm{b}}$ with spatial densities  $\LB$.\footnote{PPPs serve as a good model for the planned deployment of macro cell BSs, as verified by both empirical evidence \cite{TayDhiNov:12,DiLuGua:16} and theoretical analysis \cite{BlaKarKee:13}.}
The location of UEs is modelled as another independent PPP $\Phi_{\mathrm{u}}$ with spatial density $\LU$, where each UE associates with its closest BS for transmission. We assume the UE density is high enough that every BS has at least one UE associates with it.\footnote{The one UE per cell set up is mainly for the sake of notational simplicity. By adopting a similar approach as \cite{yang2017packet}, the analysis in this paper can be easily extended to consider more realistic scenario where multiple UEs in each cell share the same wireless channel. }
In this network, all BSs and UEs are assumed to be equipped with single antenna, and each BS transmits with constant power $P_{\mathrm{mt}}$.
In light of its spectral efficiency, we employ universal frequency reuse throughout the network, i.e., every BS transmits in the same spectrum.

We use a discrete time queuing system to model the random traffic profile. In particular, the time axis is segmented into a sequence of equal time intervals, referred to as time slots. We further assume all queuing activities, i.e., arrivals and departures, take place around the slot boundaries. Specifically, at the $m^{\text{th}}$ time slot, a potential packet departure may occur in the interval $(m^{-},m)$, and a potential packet arrival can happen in the interval $(m,m^+)$. In other words, departures occur at the moment immediately before the slot boundaries while arrivals occur at the moment immediately after the slot boundaries.
For a generic {UE} located in the cell of $x_i \in \Phi_{\mathrm{b}}$, we model its packet arrival as a Bernoulli process with rate $\xi \in [0,1]$, which represents the probability of a new arrival occurs in a slot. We further assume that each BS accumulates all the incoming packets in an infinite-size buffer for further transmission purpose.

In order to investigate the evolution of queuing dynamic, we limit the mobility of transceivers by considering a static network, i.e.,
the locations of the BSs and UEs are generated once at $(0^-, 0)$, and remained unchanged in all the following time slots.\footnote{Note that most of the pratical networks can be approximately regarded as static, since the locations of any end device cannot change drastically in a relatively short period \cite{ZhoHaeQue:16}.}

\begin{figure}[t!]
  \centering{}

    {\includegraphics[width=0.95\columnwidth]{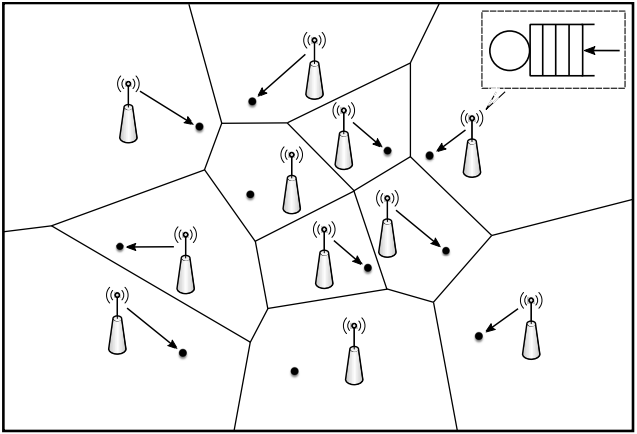}}

  \caption{A snapshot of the Poisson cellular network with temperal traffic. The coverage area of different BSs form the Voronoi cells, whereas each BS may be active or inactive, depending on its buffer status.}
  \label{fig:TrafficModel}
\end{figure}

\subsection{Propagation Channel and Failure Retransmission}
In this network, we adopt a block-fading propagation model, where the channels between any pair of antennas are assumed independent and identically distributed (i.i.d.) and quasi-static, i.e., the channel is constant during one transmission slot, and varies independently from slot to slot.
We consider all propagation channels are narrowband and affected by two attenuation components, namely small-scale Rayleigh fading with unit mean power, and large-scale path loss that follows power law.\footnote{The analysis in this paper is not necessary constrained to simple propagation model, it can be further extended to incorporate more realistic setups that include multi-slope path loss \cite{DinWanLop:16,AndZhaDur:16} and complicated fading environment \cite{ChuCotDhi:17,TriAffLia:17}. }

Affected by the random channel fading and aggregated interference, the process of packet departure does not possess a constant rate and can lead to failure packet deliverty. Retransmission is thus necessary to guarantee the packet can be correctly received. By enabling retransmission, the transmission model at each BS becomes:
During each time slot, every node with a non-empty buffer sends out a packet from the head of its queue. If the received SIR exceeds a predefined threshold, the transmission is successful and the packet can be removed from the queue; otherwise, the transmission fails and the packet remains in the buffer. We assume the feedback of each transmission, either success or fail, can be instantaneously awared by the BSs such that they are able to schedule transmission at next time slot. Moreover, for the BSs with empty buffer, they mute the transmissions to reduce power consumption and inter-cell interference.

\subsection{Signal-to-interference ratio (SIR)}
\begin{figure}[t!]
  \centering{}

    {\includegraphics[width=0.95\columnwidth]{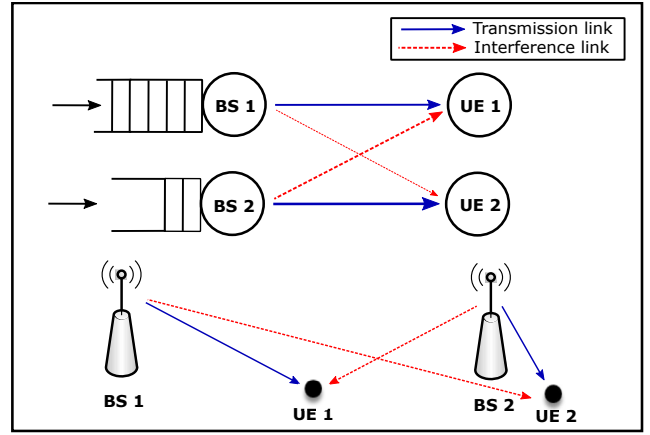}}

  \caption{Example of interacting queues with two BSs sharing the wireless channel. Due to impact from mutual interference, the service rates of BS~1 and BS~2 are different, and dynamically change with their queuing status. }
  \label{fig:Interact_Queue}
\end{figure}
By applying Slivnyak's theorem \cite{BacBla:09} to the stationary  {PPP} of {BS}, it is sufficient to focus on the {SIR} of a typical {UE} at the origin, with its tagged BS located at $x_0$.
Given the UE is receiving data at time slot $t$, the received {SIR} can be written as
\begin{align}\label{equ: SIR}
\gamma_{x_0, t} &= \frac{P_{\mathrm{mt}} h_{x_0} \Vert x_0 \Vert^{-\alpha} }{  \sum\limits_{x \in \Phi_{\mathrm{b}}\setminus x_0  } {P_{\mathrm{mt}} \zeta_{x,t} h_{x}}{ \Vert x \Vert^{-\alpha}}   },
\end{align}
where $h_x \sim \exp(1)$ denotes the small scale Rayleigh fading from BS $x$ to the origin, $\Vert \cdot \Vert$ is the Euclidean distance, $\alpha$ stands for the path loss exponent, and $\zeta_{x,t} \in \{0,1\}$ is an indicator showing whether a node located at $x \in \Phi_{\mathrm{b}}$ is transmitting at time slot $t$ ($\zeta_{x,t} = 1$) or not ($\zeta_{x,t}=0$).

It is important to note that since the spectrum is shared among BSs, the queuing status of each BS is coupled with other transmitters and hence results in interacting queue. As such, the BS active state, $\zeta_{x,t}$, is both spatial and temperal dependent, since the location affects the pathloss and further the aggregated interference, and the time changes the queue length at each node. To better illustrate this concept, Fig.~\ref{fig:Interact_Queue} gives a simple example of queuing interaction between two BSs.
Note that compared to UE~1, UE~2 has an advantage location and hence enjoys better path loss and fewer interference. Consequently, BS~2 can quickly empty its queue, and the disparity between their communication conditions results in BS~2 activates less frequently than BS~1.
Moreover, depending on whether packets appear at both BSs or not, the corresponding active durations are also different: If both transmitters have packets to send, the mutual interference will reduce the service rate and prolong the active duration of each BS individually.
On the other hand, when one transmitter becomes silent, the other one can benefit from the reduced cross-talk and speed up its queue flushing process, hence also decreases the active period.
Extending this concept to a large-scale network, we find that as the realization of PPP is irregular, there are always some BSs experience poor transmission environment, e.g., their UEs are located at the cell edge, and some others having good communication condition, e.g., their UEs are around the cell centers. In this regard, even the packet arrival rate is the same for all transmitters, the queuing status and active state can varies largely from BS to BS, and the characterization of SIR in such network is very challenging.

\section{Analysis} \label{sec:Analysis}
\newcounter{TempEqCnt}
\setcounter{equation}{\value{equation}}
\setcounter{equation}{3}
\begin{figure*}[t!]
\begin{align} \label{equ: Meta_Grl}
F^{\Phi}(u) = \frac{1}{2} - \frac{1}{\pi} \int_0^\infty \frac{1}{\omega} \mathrm{Im}\left\{ u^{-j \omega} \left[ 1 + \delta \sum_{k=1}^{\infty} \binom{j \omega}{k} \left( F^\Phi(\xi) + \int_{\xi}^1 \frac{\xi^k}{t^k}  F^\Phi(dt) \right) \mathcal{Z}(k, \delta, \theta) \right]^{-1} \right\} d\omega
\end{align}
\setcounter{equation}{\value{equation}}{}
\setcounter{equation}{4}
\centering \rule[0pt]{18cm}{0.3pt}
\end{figure*}
\setcounter{equation}{\value{TempEqCnt}}
\setcounter{equation}{1}
We now present the main technical results of the paper. In particular, we first detail the definition of conditional SIR coverage probability, and the analysis on its distribution. Then we discuss the conditions for the queuing network to be stable. After that, we give moments as well as a computationally efficient approximation for the conditional SIR coverage probability. Finally, we derive the distribution of mean delay, which involves both queuing and transmission delay.

\subsection{Conditional SIR Coverage Probability}
Since both the received signal strength and the interference at a given UE are governed by a number of stochastic processes, e.g., random spatial distribution of transmitting/receiving nodes, random packet arrivals, and queuing dynamics, the SIR in \eqref{equ: SIR} is a random variable and can only be characterized via distribution.
In this regard, conditioning on the realization of the point process $\Phi \triangleq \Phi_{\mathrm{b}} \cup \Phi_{\mathrm{u}}$, we define the conditional SIR coverage probability as follows \cite{haenggi2016meta,ZhoHaeQue:16}.
\begin{definition}
\textit{
Given the typical UE is receiving data at time slot $t$,, its conditional SIR coverage probability is defined as
\begin{align}\label{equ: ConCovProb}
\mu^\Phi_{x_0, t} &= \mathbb{P}\left( \gamma_{x_0, t} \geq \theta | \Phi \right)
\nonumber\\
&= \mathbb{E}\!\!\left[ \exp\!\left( - \theta \Vert x_0 \Vert^\alpha \!\!\!\!\! \sum_{x \in \Phi_{\mathrm{b}} \setminus x_0 } \!\!\!\! \zeta_{x, t}  \frac{ h_{x} }{ \Vert x \Vert^\alpha} \right) \Big | \Phi  \right].
\end{align}
}
\end{definition}
Note that the conditional SIR coverage probability $\mu_{x_0,t}^\Phi$ is still a random variable (as we condition on the realization of $\Phi$), which contains all the information about the UE SIR (and therefore achievable rate) distribution
across the network. Moreover, the interaction of queues is also captured by \eqref{equ: ConCovProb} via the accumulated interference.

In order to analyze the distribution of $\mu_{x_0,t}^\Phi$, we need to address two issues: $i$) due to random packet arrival and retransmission of failed deliveries, the active state, i.e., $\zeta_{x,t} = 1$, at each transmitter varies over time, and $ii$) there may exist common interferencing BSs seen by the same UE from one time slot to another, which introduces temporal correlation for the SIR coverage probability \cite{SchBetBra:12,ZhoZhaHae:14,GanHae:09}. The dynamically changing active state of BSs, together with the temperal correlation, involve memory to the queues and highly complicates the analysis.
Fortunately, when the number of transmitters asymptotically approaches infinity, a \textit{mean field} property starts to emerge in the evolution of queues, i.e.,
the interaction between queues become ``weak'' and ``global'', and the impact from aforementioned temperal and spatial correlation tends to be negligible on the employed system model \cite{BorMcDPro}.\footnote{The \textit{mean field} effect appears because of the cellular infrastruture, where interference is bounded away from the tagged transmitter. In Poisson bipolar networks, where the interaction is ``strong'' and ``local'', such approximation may not hold. }
Motivated by this fact, we make the following assumption.

\begin{assumption}
\textit{
The temporal interference correlation has a negligible effect on the transmission SIR coverage probability. Hence, we
assume the typical UE sees almost independent interference at each time slot.
}
\end{assumption}

It is now safe to assume that all UEs experience i.i.d. steady state queue distributions, and each BS activates independently, whereas the active probability depends on the specific service rate, or equivalently, the SIR coverage probability.
To faciliate analysis in the following, we introduce a simple result from queuing theory as a preliminary, which describes the active probability of a transmitter under fixed arrival and departure rates.
\begin{lemma}\label{prop:Con_MPT}
\textit{
Given the arrival rate being $\xi$, the service rate $\mu$,
the active probability at a generic BS is }
\begin{align}
\eta^a = \left \{
\begin{tabular}{cc}
\!\!\!\!\!\text{1}, & \text{if}~ $\mu \leq \xi$,   \\
\!\!\!\!  ${ \xi}/{\mu}$, & \text{if}~ $\mu >  \xi$.
\end{tabular}
\right.
\end{align}
\end{lemma}
\begin{IEEEproof}
The result is a standard conclusion from Geo/Geo/1 queue, which can be found in \cite{AteMor:04}.
\end{IEEEproof}

After all the above preparason, we are now ready to derive our main result of this paper, i.e., distribution of the conditional SIR coverage probability.
\begin{theorem}\label{thm: Dist_ConProb_ExaC}
\textit{
The cumulative distribution function of the conditional SIR coverage probability is given by the fixed-point equation \eqref{equ: Meta_Grl} on top of this page, which can be iterative solved as follows
\setcounter{equation}{\value{equation}}
\setcounter{equation}{4}
\begin{equation}
F^{\Phi}(u) = \lim_{n \rightarrow \infty} F_n^{\Phi}(u)
\end{equation}
where  $F^{\Phi}_{n}(u)$ is given by
\begin{align}
{F}_{n}^{\Phi}(x) \! &=\! \frac{1}{2} \!-\! \frac{1}{\pi}\!\! \int_{0}^\infty \!\! \frac{1}{\omega} \mathrm{Im}\!\left\{ {x^{ - j \omega}} \Big[1 + \delta \sum_{k=1}^{\infty} \! \binom{j \omega}{k} \eta_{n-1}^{(k)} \right.
\nonumber\\
& \qquad \qquad \qquad \qquad \times \left.  \mathcal{Z}\!\left( k, \delta, \theta \right) \Big]^{-1} \right\} d\omega
\end{align}
whereas $\delta = 2 / \alpha$, $\mathrm{Im}\{\cdot\}$ denotes the imaginary part of a complex number, and $\mathcal{Z}(k, \delta, \theta)$ has the form as
\begin{align}
\mathcal{Z}(k, \delta, \theta) = \frac{ (-1)^{k+1} \theta^k }{k-\delta} {}_2F_1(k, k-\delta; k-\delta+1, -\theta),
\end{align}
with ${}_2F_1(a, b; c, d)$ being the hypergeometry function \cite{AndAsk:00}, and $\eta_{n-1}^{(k)}$ is given as
\begin{align}
\eta_{n-1}^{(k)} = {F}^\Phi_{n-1}(\xi) + \int_{\xi}^1 \frac{\xi^k}{t^k} {F}^{\Phi}_{n-1}(dt).
\end{align}
In particular, when $n=1$, we have $\eta_{0}^{(k)} = \xi^k, \forall k \in \mathbb{N}$.
\setcounter{equation}{\value{equation}}
\setcounter{equation}{8}
\setcounter{equation}{\value{equation}}
}
\end{theorem}
\begin{IEEEproof}
See Appendix~\ref{appnd: Prf_Dist_ConProb_ExaC}.
\end{IEEEproof}

The expression in \eqref{equ: Meta_Grl} not only
quantifies how all the key features of a cellular network, i.e., deployment strategy, interference, and traffic profile, affect the distribution of SIR, but also illustrates how the interacting queues are affecting the SIR coverage via a fixed-point functional equation.
The function $F^\Phi(u)$ can be interpreted from two aspects:
If we regard a typical cell as a queuing system, equation~\eqref{equ: Meta_Grl} describes the distribution of the random service rate; if we look at \eqref{equ: Meta_Grl} in the view of network performance, then the CCDF, i.e.,  $1-F^\Phi(u)$, gives the level of certainty that at least $u$ fraction of UEs in the network can attain SIR threshold $\theta$.
Several numerical results based on \eqref{equ: Meta_Grl} will be shown in Section IV to provide more practical insights. In the following, we discuss the stable region, moments of the conditional SIR coverage probability, and approximation for the CDF.

\remark{ \textit{We introduce an auxiliary function as
\begin{align}\label{equ:aux}
\eta(\xi) = F^{\Phi}(\xi) + \int_{\xi}^1 \frac{\xi}{t} F^{\Phi}(dt)
\end{align}
which can be regarded as ``average'' active probability. Note that as $\xi$ grows from 0 to 1, while the value of $\eta(\xi)$ also variates accordingly, the trend is very different. As depicted in Fig.~\ref{fig:Eta_func}, we can see that, with failed packet retransmissions, the average active probability goes up on a fast-and-then-slow basis with respect to the packet arrival rate $\xi$, and this effect is especially significant in networks with high SIR detection threshold.
  }
}

\begin{figure}[t!]
  \centering{}

    {\includegraphics[width=0.95\columnwidth]{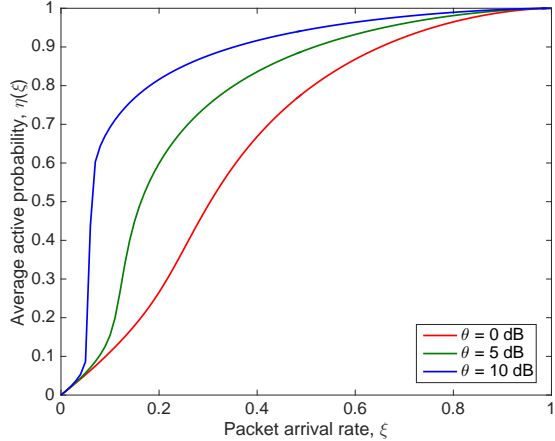}}

  \caption{Average active probability versus packet arrival rate, under different SIR detection thresholds.}
  \label{fig:Eta_func}
\end{figure}

\subsection{Stable Region}
The primal consideration in queuing systems is about stability, i.e., the critical conditions under which all the queues can remain finite length and do not explode. For an isolated system, even with random arrival and departure process, the stable region can be explicitly determined to be when average service rate is larger than the average arrival rate (Loynes theorem \cite{Loy:62}).
However, such condition cannot be directly generalized to large-scale queuing networks, as the strict stability, i.e., all queues are finite-length, is not achievable (except for the trivial case of $\xi = 0$).
Recall our discussion in Section II, the random locations of both BS and UE always result in some UEs located at poor coverage area, e.g., the cell edge, and having unbounded queue length.
In this regard, instead of requiring all the queues to be stable, a more meaningful alternative will be to maintain the fraction of unstable queues to be below certain level.
To this end, we introduce the $\varepsilon$-stable region \cite{ZhoHaeQue:16}, which gives the conditions for a network to be operated with less than $\varepsilon$ portion of saturated queues. Following \cite{ZhoHaeQue:16}, a formal definition is in the sequel.

\begin{definition}
\textit{
For any $\varepsilon \in [0, 1]$, the $\varepsilon$ - stable region $\mathcal{S}_{\varepsilon}$  and the critical arrival rate  $\xi_{\mathrm{c}}$ are defined respectively as
\begin{align}
\mathcal{S}_{\varepsilon} = \left\{ \xi \!\in\! \mathbb{R}^+ : \mathbb{P}\left\{ \lim_{T \rightarrow \infty} \frac{1}{T} \sum_{t=1}^T \mu^\Phi_{x_o} \leq \xi \right\} \leq \varepsilon \right\}
\end{align}
and
\begin{align}
\xi_{\mathrm{c}} = \sup \mathcal{S}_{\varepsilon}.
\end{align}
The network is $\varepsilon$-stable if and only if $\xi \leq \xi_{\mathrm{c}}$.
}
\end{definition}

The critical arrival rate gives an explicit stable boundary, beyond which there are more queues transfering from finite size to infinite length and the requirement of less than $\varepsilon$ portion of unstable queues cannot be guaranteed.
While an exact expression for the critical arrival rate is not available,
we can nevertheless obtain explicit conditions for the network to be $\varepsilon$-stable.
\begin{theorem}
\textit{
The sufficient condition for the network to be $\varepsilon$-stable is
\begin{align}
  \xi & \leq \xi^{\mathrm{Sc}}_{\varepsilon} \!=\! \sup\!\left\{ \xi \!\in\! \mathbb{R}^+ \!:\! \frac{1}{2} \!-\! \frac{1}{\pi}\!\! \int_{0}^\infty \!\! \frac{1}{\omega} \mathrm{Im}\!\left\{ {x^{ - j \omega}} \Big[1 \!+\! \delta \sum_{k=1}^{\infty} \! \binom{j \omega}{k}  \right. \right.
\nonumber\\
& \qquad \qquad \qquad \qquad \times \left. \left.  \mathcal{Z}\!\left( k, \delta, \theta \right) \Big]^{-1} \right\} d\omega \leq \varepsilon  \right\},
\end{align}
and the necessary condition for the network to be $\varepsilon$-stable is
\begin{align}
  \xi & \leq \xi^{\mathrm{Nc}}_{\varepsilon} \!=\! \sup\!\left\{ \xi \!\in\! \mathbb{R}^+ \!:\! \frac{1}{2} \!-\! \frac{1}{\pi}\!\! \int_{0}^\infty \!\! \frac{1}{\omega} \mathrm{Im}\!\left\{ {x^{ - j \omega}} \Big[1 \!+\! \delta \sum_{k=1}^{\infty} \! \binom{j \omega}{k} \right. \right.
\nonumber\\
& \qquad \qquad \qquad \qquad \times \left. \left. \xi^k  \mathcal{Z}\!\left( k, \delta, \theta \right) \Big]^{-1} \right\} d\omega \leq \varepsilon  \right\}.
\end{align}
}
\end{theorem}
\begin{IEEEproof}
For the sufficient condition, we consider a dominant system, where all the nodes keep transmitting irrespect of their buffer status (for a node with empty buffer, it transmits ``dummy'' packets). As such, the interference in dominant system is larger than the actual ones, and the result can be attained via having $\xi = 1$ in \eqref{equ: Meta_Grl}.

For the necessary condition, we consider a favorable system, where each node transmits without retransmission, i.e., even there is transmission failure, the transmitter simplily ignores the failure and withdraw the failed packets. In this scenario, the interference is smaller than that of the actual one, whereas we can obtain the result as making $\eta_k = \xi^k$.
\end{IEEEproof}

From the above results, we note that the actual value of the critical arrival rate falls in the interval of $[\xi^{\mathrm{Sc}}, \xi^{\mathrm{Nc}}]$. Using the ergodic property of PPP, an approximation for the $\varepsilon$-stable region is given in the sequel.
\begin{corollary}
\textit{
The  $\varepsilon$-stable region of the network is approximated as follows
\begin{align}
\xi \leq \xi^{\mathrm{A}}_{\mathrm{c, \varepsilon}} = \sup\left\{ \xi \in \mathbb{R}^+ : F^\Phi(\xi) \leq \varepsilon \right\}.
\end{align}
}
\end{corollary}
\begin{IEEEproof}
Note that due to stationary and ergodicity, the ensemble average obtained by averaging over the point process equals the spatial averages obtained by averaging an arbitrary realization of PPP over a large region, i.e.,
\begin{align}
\mathbb{P}^{x_0}\left( \lim_{T \rightarrow \infty} \frac{1}{T} \sum_{t=1}^T \mu_{x_0, t}^\Phi \leq \xi \right) = F^\Phi(\xi),
\end{align}
and the result follows according to the definition.
\end{IEEEproof}

\remark{\textit{
  With the notion of stability, the function $\eta(\xi)$ defined in \eqref{equ:aux} can be interpreted as: On average, there are $F^\Phi(\xi)$ portion of UEs in the network with infinity queue size and keep transmitting, while the rest $1-F^\Phi(\xi)$ portion of UEs maintain stable queues, and active independently with probability $\int_{\xi}^1 \xi / t F^\Phi(dt)$.
}}

\subsection{Moments}
Based on the CDF of the conditional SIR coverage probability, in this part we give the corresponding moments, which can faciliate the assessment of network performance.
\begin{theorem}\label{thm: moment}
\textit{
The $m$-th moment of the conditional SIR coverage probability is given by
\begin{align}
M_{m} =  \frac{1}{1 + \delta \! \sum_{k=1}^m \! \binom{m}{k} \eta^{(k)} \! \mathcal{Z}(k, \delta, \theta)}  ,
\end{align}
where $\eta^{(k)}$ is given as
\begin{align}
\eta^{(k)} = {F}^\Phi( \xi) + \int_{\xi}^1 \left(\frac{\xi}{t}\right)^k {F}^{\Phi}(dt)
\end{align}
with $F^\Phi(u)$ giving in \eqref{equ: Meta_Grl}.
In particular, when $m=1$, we have the standard SIR coverage probability given as
\begin{align}
&\mathbb{P}\left( \gamma_{x_0} > \theta \right) = \frac{1}{1 + \delta \eta \mathcal{Z}(1, \delta, \theta) }.
\end{align}
}
\end{theorem}
\begin{IEEEproof}
See Appendix~\ref{appnd: Prf_moment}.
\end{IEEEproof}

Because of its important role in network performance assessment, we further provide bounds as well as an approximation for the SIR coverage probability to attain better insights.
\begin{corollary}
\textit{The SIR coverage probability can be respectively bounded by the following
\begin{align}
\frac{1}{1 + \delta \mathcal{Z}(1, \delta, \theta) } < \mathbb{P}(\gamma_{x_0} > \theta) < \frac{1}{1 + \delta \xi \mathcal{Z}(1, \delta, \theta) }
\end{align}
and when $\xi \ll 1$, the SIR coverage probability can be approximated as follows
\begin{align}
\mathbb{P}(\gamma_{x_0} > \theta) & \approx 1 - \xi \delta \mathcal{Z}(1, \delta, \theta)
\nonumber\\
 & \stackrel{(\alpha=4)}{=} 1 - \xi \delta \sqrt{\theta} \arctan(\sqrt{\theta}).
\end{align}
}
\end{corollary}
\begin{IEEEproof}
The upper and lower bounds can be obtained via similar approach as in the proof of Theorem~2, i.e., by considering a favorable system without retransmission and a dominant system that keeps transmitting, respectively.

For the approximation, we notice that $F^\Phi(\xi) \rightarrow 0$ as $\xi \rightarrow 0$. Hence, we assume all the queues are stable in the regime with very light traffic load, and approximate the active probability using the mean, i.e., $\xi / \mathbb{P}(\gamma_{x_0} > \theta)$ \cite{yang2017packet}. The result then follows from solving the following equation
\begin{align}
\mathbb{P}(\gamma_{x_0} > \theta) = \frac{1}{1 +  \frac{\delta \xi}{\mathbb{P}(\gamma_{x_0} > \theta)} \mathcal{Z}(1, \delta, \theta) }.
\end{align}
\end{IEEEproof}

\remark{
\textit{
  Note that the gap between the upper and lower bounds decreases with the increment of the packet arrival rate $\xi$, which is consistent with the conclusion in \cite{ZhoQueGe:16}. Furthermore, in the light traffic regime, the outage probability, $1-\mathbb{P}(\gamma_{x_0} > \theta)$, increases linearly with respect to the packet arrival rate $\xi$, which demonstrates the significant impact of traffic profile on the system performance.
}
}

\subsection{Approximation}
Motivated by the fact that $F^\Phi(u)$ is supported on the interval $[0, 1]$, we content ourselves in this part by approximating the function $F^\Phi(u)$ via a Beta distribution to reduce the computational complexity \cite{haenggi2016meta}.
A formal operation is stated in the sequel.

\begin{corollary}\label{Cor:Approximation}
\textit{The probability distribution function (pdf) of $F^{\Phi}_{n}(x)$ in Theorem 1 can be tightly approximated via the following
\begin{align}
f_{X_n}(x) = \frac{x^{\frac{\mu_n (\beta_n + 1) - 1}{1 - \mu_n}} (1-x)^{\beta_n - 1} }{B(\mu_n \beta_n/(1-\mu_n), \beta_n )}
\end{align}
where $B(a, b)$ denotes the Beta function, $\mu_n$ and $\beta_n$ are respectively given as
\begin{align} \label{equ:mu_n}
\mu_n &= M_1^{(n)}, \\ \label{equ:beta_n}
\beta_n &= \frac{(\mu_n - M_2^{(n)}) ( 1 - \mu_n )}{M_2^{(n)} - \mu_n^2 }
\end{align}
where $M_m^{(n)}$ can be written as
\begin{align}
M_m^{(n)} = \frac{1}{ 1 +  \delta \! \sum_{k=1}^m \!\! \binom{m}{k}  \hat{\eta}^{(k)}_{n-1} \! \mathcal{Z}(k, \delta, \theta) },
\end{align}
with $\hat{\eta}^{(k)}_{n-1}$ being
\begin{align}
\hat{\eta}^{(k)}_{n-1} = \int_0^{\xi} \!\! f_{X_{n-1}}(t) dt +\! \int_{\xi}^1 \! \frac{\xi^k}{t^k} f_{X_{n-1}}(t) dt.
\end{align}
When $n=1$, we have $\hat{\eta}_{0}^{(k)} = \xi^k, \forall k \in \mathbb{N}$.
}
\end{corollary}
\begin{IEEEproof}
At each step of the iteration in Theorem~\ref{thm: Dist_ConProb_ExaC}, the function $F^\Phi_n(u)$ is supported on $[0, 1]$. As such, by respectively matching the mean and variance to a Beta distribution $B(a_n, b_n)$, it yields
\begin{align}
&\frac{a_n}{a_n + b_n} = M_1^{(n)},\\
&\frac{a_n b_n}{ (a_n + b_n)^2 (a_n + b_n + 1) } = M^{(n)}_2 - \left[ M_1^{(n)} \right]^2
\end{align}
and the result follows by solving the above system equations.
\end{IEEEproof}

The accuracy of Corollary~\ref{Cor:Approximation} will be verified in Fig.~\ref{fig:Beta_Apprx}.
\subsection{Mean Delay and Analysis}
In the context of wireless networks, delay is an important factor that determines the Quality of Service (QoS) \cite{ZhoHaeZhe:16,ZhoQueGe:16}.
As shown in Fig.~\ref{fig:Delay_Illustration}, the random arrival of packets and interference-limited channel will inevitably incur waiting and retransmission in the packet delivery process, thus induce multiple slots for one successful delivery. Besides, since the number of required time slots differs among each packet, we can only quantify the delay in average manner. In its accordance, the following definition formalizes the notion of mean day.

\begin{definition}
\textit{
Let $A_x(T)$ be the number of packets arrived at a typical transmitter $x$ within period $[0,T]$, and $D_{i, x}$ be the number of time slots between the arrival of the $i$-th packet and its successful delivery. The mean delay is defined as
\begin{align}\label{equ:Gen_Delay}
\mathbf{D}^\Phi_x \triangleq    \lim\limits_{T \rightarrow \infty} \frac{\sum_{i=1}^{A_x(T)} D_{i, x} }{A_x(T)}.
\end{align}
}
\end{definition}

Note that $D_{i,x}$ in \eqref{equ:Gen_Delay} represents the number of time slots required to successfully deliver the $i$-th packet, and its value is affected by: $(i)$ queueing delay, caused by other accumulated unsent packets, and $(ii)$ transmission, or equivalently local, delay, due to link failure and retransmission \cite{Hae:13,ZhaQueHua:16,ZhaQueKou:16}.
By averaging over all time slots, \eqref{equ:Gen_Delay} provides information on the average number of slots to successfully deliver a packet.
Furthermore, using results from queuing theory and stochastic geometry, we can characterize its distribution with the following expression.
\begin{figure}[t!]
  \centering{}

    {\includegraphics[width=0.95\columnwidth]{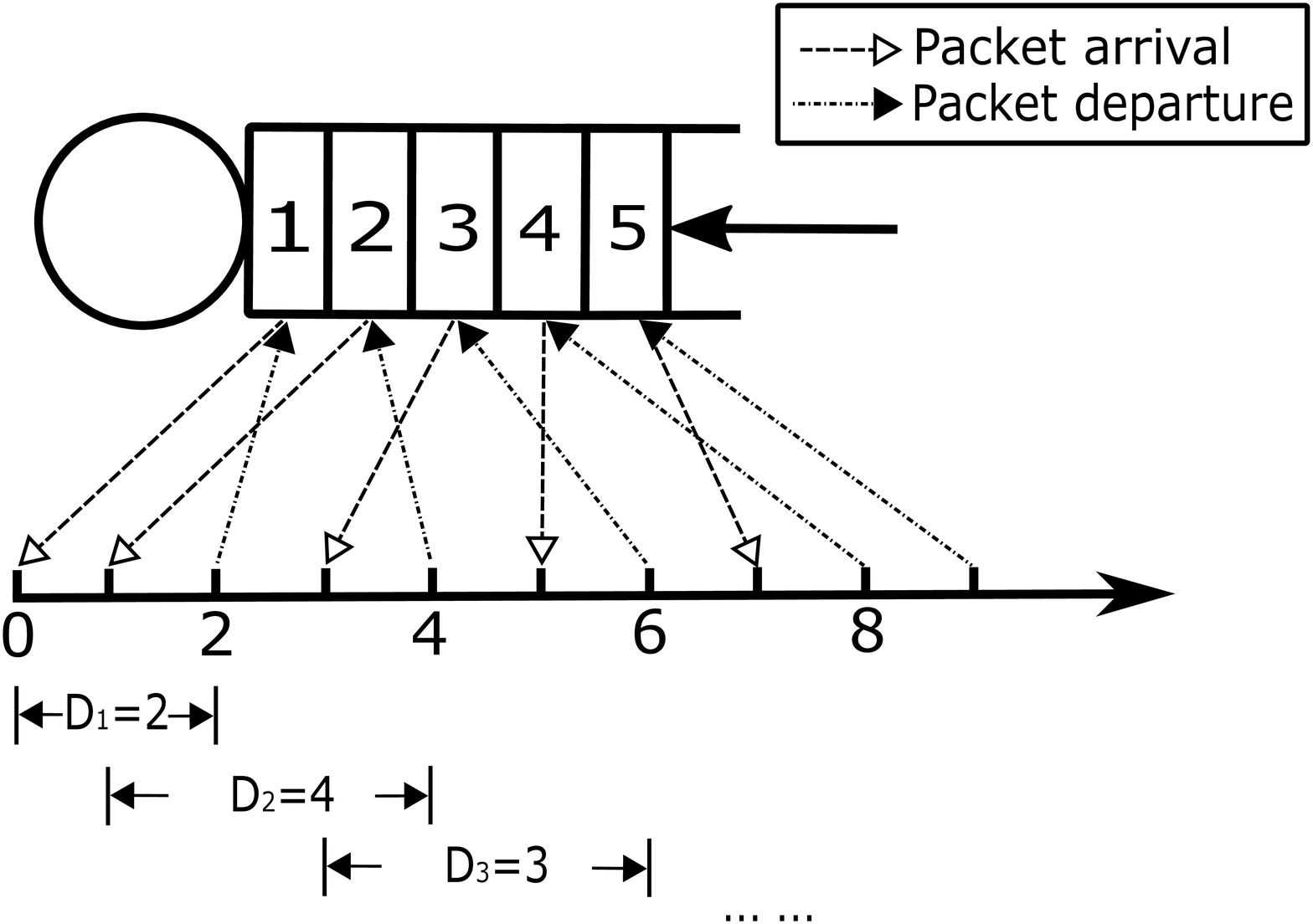}}

  \caption{Illustration of delay in queuing system. The number of required slots to deliver each packet varies due to different queuing and retransmission results. }
  \label{fig:Delay_Illustration}
\end{figure}

\begin{theorem}
\textit{
The CDF of mean delay at a typical UE is given as
\begin{align}\label{equ: Delay_Dist}
\mathbb{P}\left( \mathbf{D}_{x_0}^\Phi \leq T \right) &= {1 - F^{\Phi} \left( \frac{1-\xi}{T} + \xi \right)}
\nonumber\\
&\approx \lim_{n \rightarrow \infty} \int_{\xi + \frac{1-\xi}{T}}^{1} \frac{x^{\frac{\mu_n (\beta_n + 1) - 1}{1 - \mu_n}} (1-x)^{\beta_n - 1} }{B(\mu_n \beta_n/(1-\mu_n), \beta_n )} dx
\end{align}
where $F^\Phi(u)$ is given by \eqref{equ: Meta_Grl}, and $\mu_n$ and $\beta_n$ are respectively given in \eqref{equ:mu_n} and \eqref{equ:beta_n}.
}
\end{theorem}
\begin{IEEEproof}
Without loss of generality, we focus on the typical UE. Given the service rate, i.e., the conditional success probability, $\mu^\Phi_{x_0}$, we have the conditional mean delay being \cite{ZhoQueGe:16}
\begin{align}
\mathbf{D}_{x_0}^\Phi = \left \{
\begin{tabular}{cc}
$\frac{1-\xi}{ \mu^\Phi_{x_0} - \xi}$, & if ~ $\mu^\Phi_{x_0} > \xi$,   \\
~\\
+$\infty$, &  if ~ $\mu^\Phi_{x_0} \leq \xi$.
\end{tabular}
\right.
\end{align}
The distribution of mean delay can then be computed as
\begin{align}
\mathbb{P}\left( \mathbf{D}_{x_0}^\Phi \leq T \right) &\stackrel{(a)}{=} \mathbb{P}\left( \mathbf{D}_{x_0}^\Phi \leq T | \mu^\Phi_x > \xi \right) \mathbb{P}\left( \mu^\Phi_x > \xi \right)
\nonumber\\
&+ \mathbb{P}\left( \mathbf{D}_{x_0}^\Phi \leq T | \mu^\Phi_x \leq \xi \right) \mathbb{P}\left( \mu^\Phi_x \leq \xi \right)
\nonumber\\
&\stackrel{(b)}{=} \mathbb{P}\left( \mathbf{D}_{x_0}^\Phi \leq T, \mu^\Phi_x > \xi \right)
\nonumber\\
&= \mathbb{P}\left( \frac{1-\xi}{T} + \xi \leq \mu^\Phi_x \right)
\end{align}
where ($a$) is by the law of total probability, and ($b$) follows by noticing $\mathbb{P}(\mathbf{D}_{x_0}^\Phi \leq T | \mu^\Phi_x \leq \xi ) = 0$. We then obtain the result by using \eqref{equ: Meta_Grl} and Corollary~3 to the above.
\end{IEEEproof}

The accuracy of Theorem~4 will be verified in Fig.~\ref{fig:VerF_Delay}. Moreover, two observations immediately follows from \eqref{equ: Delay_Dist}.

\textbf{Observation 1:} \textit{When $T=1$, we have $\mathbb{P}\left( \mathbf{D}_{x_0}^\Phi \leq T \right) = 0$, which states that the UEs who are able to success without retransmission form a probabilistic null set in large queuing networks.}

\textbf{Observation 2:} \textit{As $T \gg 1$, the CCDF of mean delay can be approximated as, $\mathbb{P}( \mathbf{D}_{x_0}^\Phi > T ) \approx F^\Phi(\xi) + 1/T$, which implies the distribution is heavy tail. In addition, when $T \rightarrow \infty$, we have the delay outage, i.e., $1-\mathbb{P}(\mathbf{D}_{x_0}^\Phi \leq T)$ converges to $F^\Phi(\xi)$, which represents the fraction of UEs that are experiencing unstable queues, and it is in line with our discussion in Section-III-B.
}


\section{Numerical Results and Simulations} \label{sec:NumAnal}

In this section, we validate the accuracy of our analysis through simulations, and explore the impact of traffic condition on network performance from several aspects. During each simulation run, the BSs and UEs are realized over a 100 $\text{km}^2$ area via independent PPPs. Packets arrive at each node according to independent Bernoulli process. We average over 10,000 realizations and collect the statistic from each cell to finally calculate the SIR coverage probability.
Unless differently specified, we use the following parameters for path loss exponent, BS density, and packet arrival rate, respectively: $\alpha = 3.8$, $\LB = 10^{-4}~\text{BS}/\text{km}^2$, and $\xi = 0.3~\text{packet}/\text{slot}$.

\begin{figure*}[t!]
  \centering

  \subfigure[\label{fig:1a}]{\includegraphics[width=0.95\columnwidth]{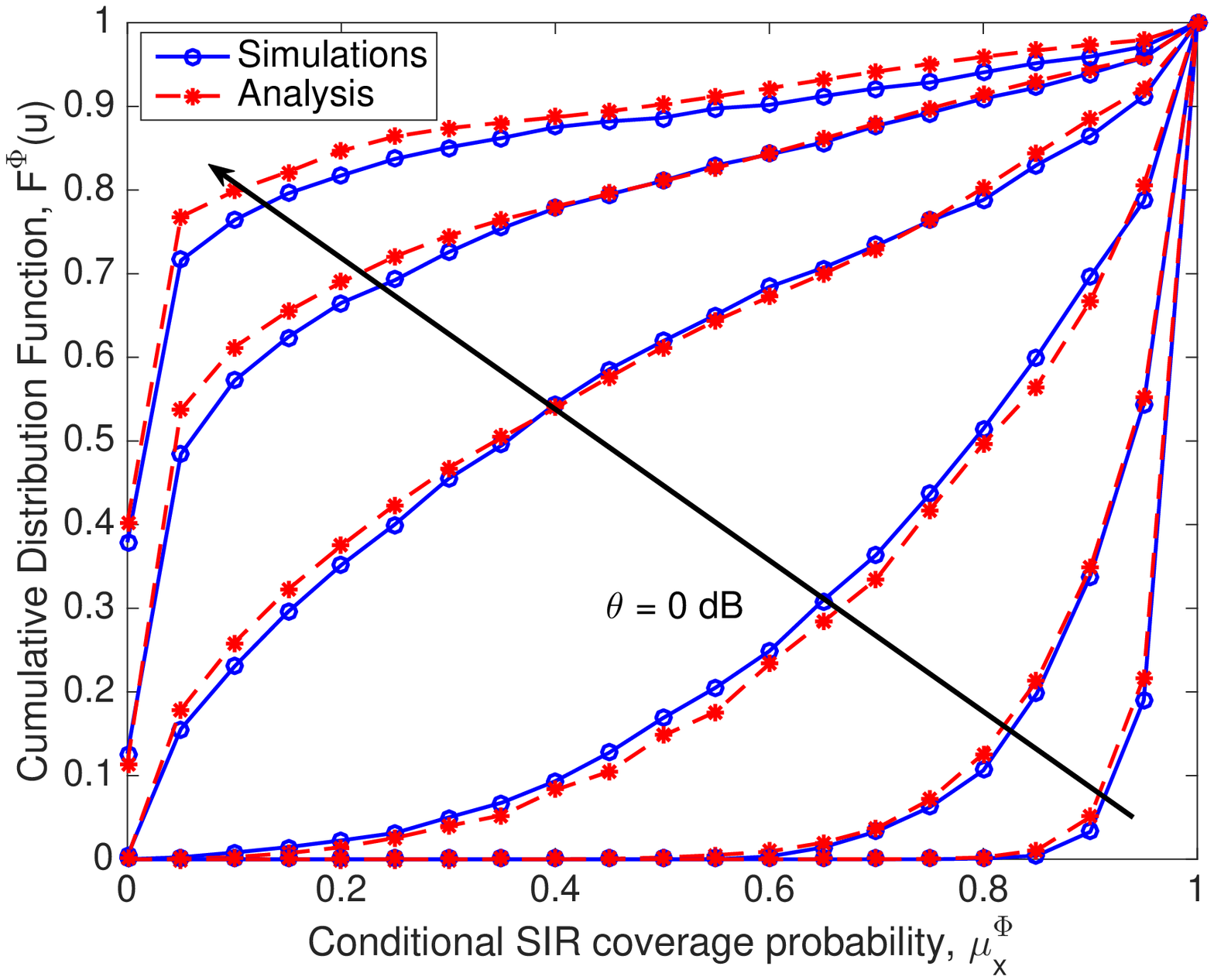}}~
  \subfigure[\label{fig:1b}]{\includegraphics[width=0.95\columnwidth]{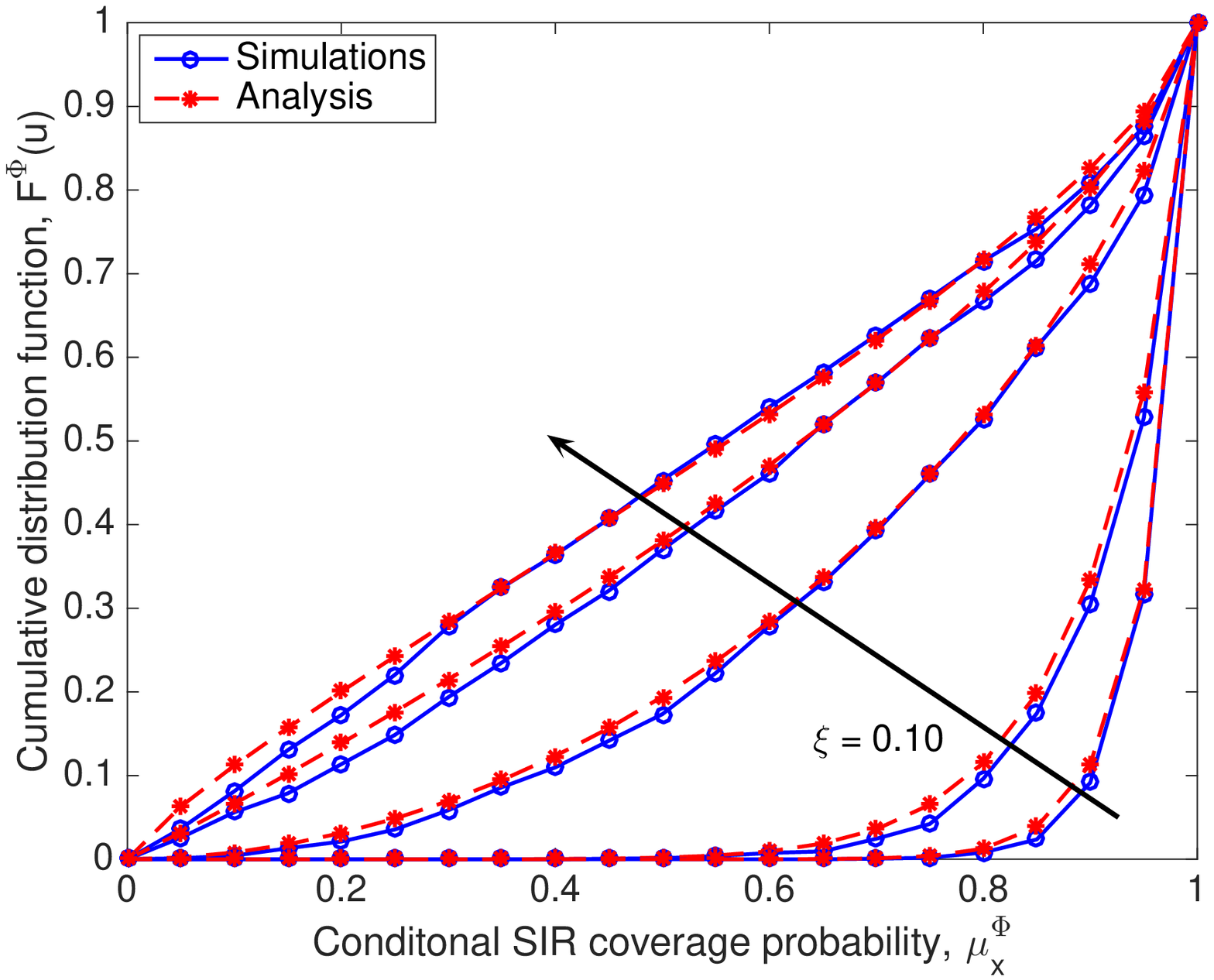}}
  \caption{Simulation versus analysis: CDF of conditional success probability. In Fig. (a), we fix the packet arrival rate to be $\xi = 0.3$, and varies the SIR threshold as $\theta = -10, -5, 0, 5, 10, 15$ dB. In Fig. (b), we fix the SIR threshold to be $\theta = 0$ dB, and varies the arrival rate as $\xi = 0.05, 0.1, 0.3, 0.5, 0.8$. }
  \label{fig:VerF_MeTa}
\end{figure*}
In Fig.~\ref{fig:VerF_MeTa}, we compare the simulated CDF of conditional SIR coverage probability to the analysis proposed in Theorem~\ref{thm: Dist_ConProb_ExaC}, for various values of SIR detection threshold $\theta$ and packet arrival rate $\xi$.
First, the results show a close match for all values of $\theta$ and $\xi$, which validate the our mathematical framework.
Next, note that from Fig.~\ref{fig:1a} we can quantify the level of confidence about how much fraction of UEs are able to achieve the targeted SIR at different thresholds, it is also coined as the reliability of the network \cite{haenggi2016meta}.
For instance, when $\xi=0.3$, the probability that 80\% of UEs in the network can attain -5~dB SIR is 0.85, showing that with high possibility the majority of UEs can have their packets correctly decoded without retransmission. Such probability can drop to 0.08 if the SIR detection threshold increases to 5~dB, indicating that in large-scale queuing networks, it is very unlikely to maintain the majority of UEs at high SIR.
On the other hand, from Fig.~\ref{fig:1b} we observe that the conditional SIR coverage probability monotonically decreases with the growth of packet arrival rate, due to the fact that more and more BSs are activated because of their non-empty buffers and thus impose additional inter-cell interference.
It is also worthwhile to note that the increment of traffic load defects the network SIR coverage in a non-linear manner, whereas the degradation of SIR coverage is more severe as traffic load goes from light ($\xi = 0.05$) to medium ($\xi = 0.3$), and the decreasing trend slows down as the network load further increases to heavy traffic regime ($\xi = 0.8$).
The reason comes from the composite effect of temporal traffic. In light traffic condition, as packet arrival rate goes up, the increased traffic load not only wakes up more BSs, but also brings in more accumulated packets at the buffer.
Together with the reduced service rate, the active duration of transmitters is extended, which in turns defect the SIR across the network.
In the heavy traffic regime, as most of the queues are already saturated, the additional active cells cannot largely change the interference, and thus the SIR coverage probability descent is leveled off. Such explanation also goes to the observation in Remark~1.

Taking a closer look between Fig.~\ref{fig:1a} and Fig.~\ref{fig:1b}, we notice that under the same cellular configuration, the conditional SIR coverage probability with $\theta = 0$ dB is larger than that with $\theta = 5$ dB, even their respective packet arrival rates are $\xi = 0.8$ and $\xi = 0.3$. This observation indicates that packets with smaller detection threshold, i.e., short packets, are more preferable to the network.
The reason comes from the fact that higher SIR threshold not merely increases the possibility of failure in decoding process, but triggers more retransmissions thus prolong the active period of interferers, which further reduces the coverage probability.

\begin{figure}[t!]
  \centering{}

    {\includegraphics[width=0.95\columnwidth]{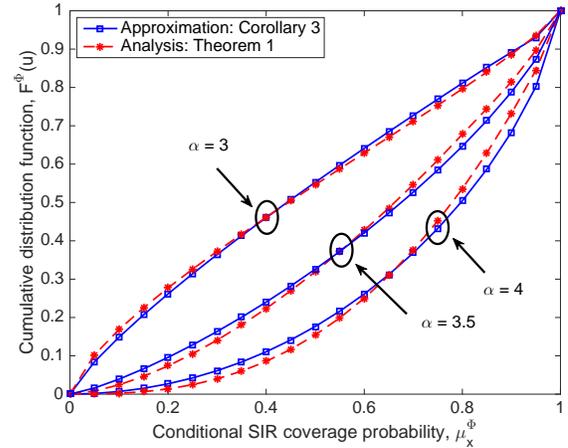}}

  \caption{CDF of conditional SIR coverage probability (Theorem~\ref{thm: Dist_ConProb_ExaC}) and the approximation (Corollary~3). Parameters are set as: SIR detection threshold $\theta = 0$~dB, and packet arrival rate $\xi = 0.3$.}
  \label{fig:Beta_Apprx}
\end{figure}

Fig.~\ref{fig:Beta_Apprx} compares the CDF of conditional SIR coverage probability to its approximation in Corollary~3. We can see that the approximation matches well with the analysis under different path loss exponents, thus confirms its accuracy. We further observe that smaller path loss exponent reduces the SIR coverage, which is due to the fact that while smaller $\alpha$ enhances the received signal power, it nevertheless also results in higher inter-cell interference, and the latter subpass the power gain and hence defacts the SIR.

\begin{figure}[t!]
  \centering{}

    {\includegraphics[width=0.95\columnwidth]{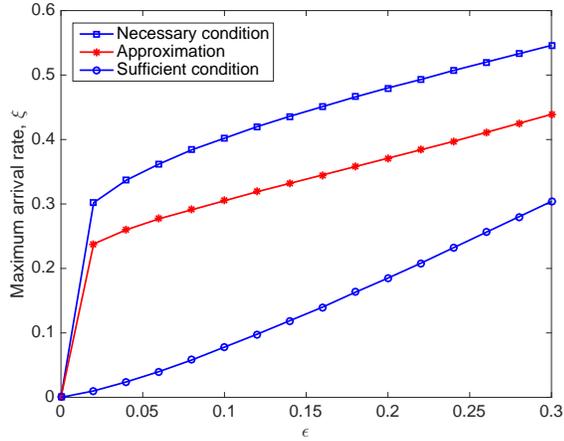}}

  \caption{Comparison of necessary and sufficient conditions.}
  \label{fig:VerF_StRgn}
\end{figure}

Fig.~\ref{fig:VerF_StRgn} plots the maximum arrival rates per sufficient and necessary conditions as functions of $\varepsilon$. The figure shows a large difference between the necessary and sufficient conditions, while the approximation locates in between, hence confirms the necessity for better SIR characterization in traffic network. It also reveals that a slight change of arrival rate can largely impact the stable region, i.e., the network stability is vulnerable to traffic condition.

\begin{figure}[t!]
  \centering{}

    {\includegraphics[width=0.95\columnwidth]{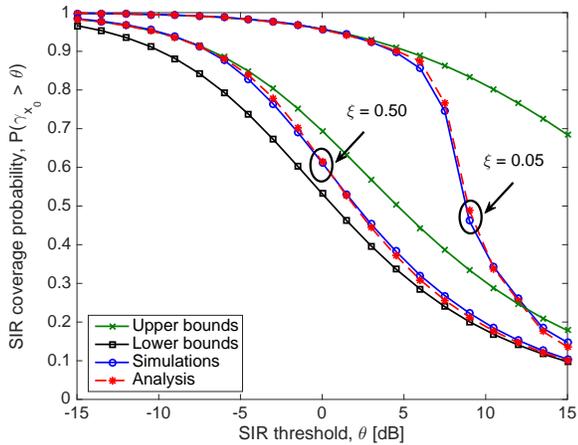}}

  \caption{Success probability versus SIR threshold $\theta$.}
  \label{fig:VerF_SINR}
\end{figure}

Fig.~\ref{fig:VerF_SINR} depicts the standard SIR coverage probability as a function of SIR detection threshold $\theta$, under different packet arrival rates. We first note the close match between the analysis and the simulation, which validates the accuracy of Theorem~\ref{thm: moment}.
More importantly, the figure confirms that the traffic profile plays a crucial role in the SIR coverage probability. For instance, to maintain a 90\% coverage probability, the required SIR detection threshold can differ more than 10 dB in light traffic ($\xi = 0.05$) and heavy traffic ($\xi = 0.5$) regimes, and this gap will be larger as the packet arrival rate keeps increasing.
Fig.~\ref{fig:VerF_SINR} also reveals that the upper and lower bounds are not tight, whereas even in heavy traffic regime the difference between the two bounds is around 5 dB, and this gap increases dramatically in light traffic condition. Moreover, the bounds fail to capture the trend of coverage probability variates according to the detection threshold $\theta$. As what we have discussed in Remark~1, both $\xi$ and $\theta$ significantly affect the active probability, hence, simply approximating the transmitting BSs to a thinned point process cannot provide desire results and hence validate the importance of the traffic-aware analysis.

In Fig.~\ref{fig:F_95_likely} we plot the cell-edge coverage probability, i.e., $1 - F^\Phi(0.95)$, as a function of packet arrival rate $\xi$. This quantity represents the performance of the ``5\% UEs'', i.e., the UEs in the bottom 5th percentile in terms of performance, and is particularly interested to operators \cite{AndBuzCho:14,haenggi2016meta}.
We can see that the cell-edge coverage probability is vulnerable to traffic fluctuation. Even in the case with very small SIR detection threshold, e.g., $\theta = -10$~dB, the cell-edge coverage probability can drop by half as the packet arrival rate changes from low to high, and this defection is more severe in scenarios with high SIR thresholds.
As such, it is critical to employ more advanced technology to boost up the SIR performance at the cell edge \cite{YanGerQue:17}.

\begin{figure}[t!]
  \centering{}

    {\includegraphics[width=0.95\columnwidth]{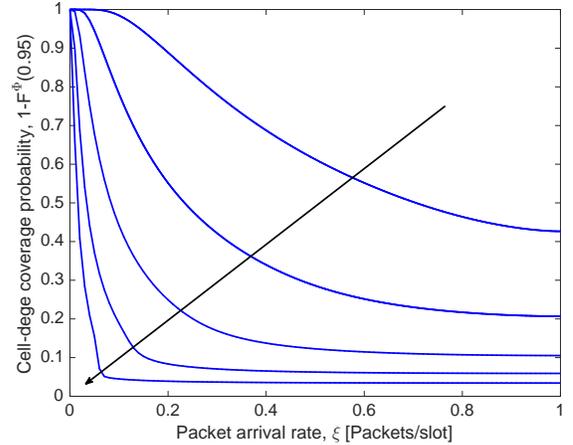}}

  \caption{95\%-likely rate versus packet arrival rate $\xi$, for various SIR thresholds $\theta = -10, -5, 0, 5, 10$ dB.}
  \label{fig:F_95_likely}
\end{figure}

Fig.~\ref{fig:VerF_Delay} compares the CDF of mean delay given in Theorem~4 to the values obtained from simulations. The figure shows that simulation results agree well with the analytical values.
Moreover, Fig.~\ref{fig:VerF_Delay} confirms the observation in Section III by showing a heavy tail behavior in the distribution of mean delay.
Fig.~\ref{fig:VerF_Delay} also reveals that the mean delay is very sensitive to the variation of traffic load, e.g., the delay outage is round 0.01 when $\xi=0.2$, but this value climbs to around 0.25 when the packet arrival rate doubles, i.e., $\xi=0.4$.
It is becuase higher the traffic load, on one hand incurs more retransmissions via increased interference, on the other, also prolongs waiting time of each packet since the additional incomings quickly occupy all available buffers. This composite effect significantly affects the mean delay.

\begin{figure}[t!]
  \centering{}

    {\includegraphics[width=0.95\columnwidth]{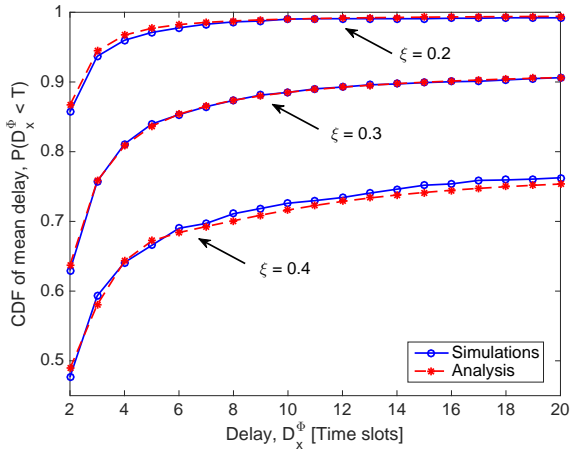}}

  \caption{Simulation versus analysis: Mean delay distribution.}
  \label{fig:VerF_Delay}
\end{figure}

\section{Conclusion}\label{sec:conclusions}
In this paper, we introduced an analytical toolset to evaluate the impact of temporal traffic on the performance of cellular networks. We used a general model that accounts for key features from both spatial and temporal domain, including the channel fading, path loss, network topology, traffic profile, and queuing evolution.
By exploiting queuing theory and stochastic geometry, we obtain the SIR distribution through a fixed-point functional equation, and validated its accuracy by simulation.
Our results confirmed that temporal traffic profile can largely affect the network SIR performance. In particular, it showed that under the same configuration, when traffic condition changes from light to heavy, the corresponding SIR requirement can differ by more than 10~dB for the network to maintain coverage. Moreover, the SIR coverage probability varies largely with traffic fluctuation in the sub-medium load regime, whereas in scenario with very light traffic load, the SIR outage probability increases linearly with the packet arrival rate. In addition, the mean delay, as well as coverage probability of cell-edge UEs are vulnerable to the traffic fluctuation, thus confirms its appeal for traffic-aware communication technologies.

The derivation of SIR distribution as a tractable form of system parameters opens various areas to gain further design insights.
On one hand, the framework can be extended to adopt more sophisticated point process, e.g., the Poisson cluster point process \cite{GanHae:09cluster}, or determinantal point processes \cite{LiBacDhi:15}. On the other hand, the analysis can be applied to investigate the design of different wireless technologies. For example, the traffic scheduling problem in large-scale wireless networks \cite{ZhoQueGe:16}, or the resource allocation problem in Dynamic TDD system \cite{SheKhoEri:12}. Analyzing impact of temporal traffic on the performance of Massive-MIMO system is also a concrete direction to investigate in the future.

\begin{appendix}
\subsection{Proof of Theorem~\ref{thm: Dist_ConProb_ExaC}} \label{appnd: Prf_Dist_ConProb_ExaC}
To faciliate the presentation, we denote $\mathcal{F}_n$ as the $\sigma$-algebra that contains all the information about queuing status of every node $x \in \Phi_{\mathrm{b}}$ up to time $t=n$. We further introduce $Y^{\Phi}_{x,n}$ and $q_{x,n}$ to denote $Y^{\Phi}_{x,n} = \ln \mathbb{P}( \gamma^\Phi_{x, n} > \theta | \Phi )$ and $q_{x,n} = \mathbb{P}(\zeta_{x, n} = 1)$, respectively.

As such, the $\sigma$-algebra $\{\mathcal{F}_n\}_{n=0}^\infty$ forms a filtration, where $\mathcal{F}_{n-1} \subset \mathcal{F}_n$. At slot $t=0$, every transmitter has one packet arrival with probability $\xi$, and actives with probability $\xi$, i.e., $q_{x,0} = \xi$, $\forall x \in \Phi_{\mathrm{b}}$. We can thus compute the moment generation function of $Y^{\Phi}_{x,0}$ at a generic BS $x \in \Phi_{\mathrm{b}}$, via the following
\begin{align}
&M_{Y^{\Phi}_{x,0}}(s) = \mathbb{E}\left[ \exp\left( s Y^{\Phi}_{x, 0} \right) \right]
\nonumber\\
&= \mathbb{E}\left[ \prod_{z \in \Phi \setminus x } \left( \frac{\xi}{1+\theta \Vert x \Vert^{\alpha}/ \Vert z \Vert^{\alpha} } + 1 - \xi \right)^{s} \right]
\nonumber\\
&\stackrel{(a)}{=} \exp\!\left(\! - \LB \!\! \int_{B^c(0, \Vert x \Vert)}^\infty\! \left[ 1 \!-\! \left( \frac{\xi}{ 1 \!+\! \theta \Vert x_0 \Vert^\alpha / \Vert x \Vert^\alpha  } \!+\! 1 \!-\! \xi \right)^s \right]\! dx \right)
\nonumber\\
&\stackrel{(b)}{=} \int_{0}^\infty 2 \pi \LB r \exp\left( - \LB \pi r^2 \right)
\nonumber\\
&\times \exp\!\left(\! - 2 \pi \LB r^2 \! \sum_{k=1}^\infty \!\binom{s}{k} \! (-1)^{k+1} \!\!\! \int_{1}^\infty \!\!\! \left( \frac{\xi}{1 \!+\! v^{\alpha/2}/\theta } \right)^k \!\!\! v dv \right) dr
\nonumber\\
&\stackrel{(c)}{=} \Big[1 + \delta \sum_{k=1}^{\infty} \! \binom{j \omega}{k} \xi^{k}  \mathcal{Z}\!\left( k, \delta, \theta \right) \Big]^{-1}
\end{align}
where ($a$) follows from the probability generating functional of PPP, ($b$) comes from polar coordinate transform and deconditioning $\Vert x \Vert$ with its pdf, which follows a Rayleigh distribution as $f_R(r) = 2 \pi \LB r e^{- \LB \pi r^2}$, and ($c$) is by algebraic operation.
By the Gil-Pelaez theorem \cite{Gil}, we have the CDF of $\mu^\Phi_{x, 0}$ given as
\begin{align}
F^\Phi_{0}(u) &= \mathbb{P}\left( \mathbb{P}\left( \gamma_{x, 0} > \theta | \Phi \right) < u \right)
= \mathbb{P}\left( Y^\Phi_{x,0} < \ln u \right)
\nonumber\\
&= \frac{1}{2} - \frac{1}{\pi} \int_{0}^{\infty} \frac{1}{\omega} \mathrm{Im}\left\{ u^{-j \omega} M_{Y^\Phi_{x, 0}}(j \omega) \right\} d\omega.
\end{align}

Next, consider all the queues have evolved to time $t=n$, and we have obtained the CDF of $\mu^\Phi_{x,n-1}$ as $\mathbb{P}(\mu^\Phi_{x,n-1} < u ) = F^\Phi_{n-1}(u)$, the moment generating function of $Y^\Phi_{x, n}$ can be calculated as
\begin{align}\label{equ: M_Yn}
&M_{Y^\Phi_{x,n}}\!(s) =  \mathbb{E}\left[ \mathbb{P}\left( \gamma_{x, n} > \theta | \Phi \right)^s \right]
\nonumber\\
&= \mathbb{E}\!\left[ \mathbb{E}\left[ \mathbb{P}\left( \gamma_{x, n} > \theta | \Phi \right)^s  \big| \mathcal{F}_{n-1} \right] \right]
\nonumber\\
&= \mathbb{E}\!\left[ \mathbb{E} \! \left[ \prod_{z \in \Phi \setminus x } \!\!\! \left( \frac{q_{z, n}}{1 \!+\! \theta \Vert x \Vert^{\alpha}/ \Vert z \Vert^{\alpha} } \!+\! 1 \!-\! q_{z, n} \right)^{s}  \Bigg |  \mathcal{F}_{n-1} \right] \right]
\nonumber\\
&= \int_{0}^\infty \!\!\! 2 \pi \LB r \exp\left( - \LB \pi r^2 \right) \exp\!\left(\! - 2 \pi \LB r^2 \! \sum_{k=1}^\infty \!\binom{s}{k} (-1)^{k+1} \right.
\nonumber\\
&\times \left. \mathbb{E}\left[ \int_{1}^\infty \!\!\! \left( \frac{q_{x,n}}{1 \!+\! v^{\alpha/2}/\theta } \right)^k \!\!\! v dv \Bigg| \mathcal{F}_{n-1} \right] \right) dr
\nonumber\\
&= \Big[1 + \delta \sum_{k=1}^{\infty} \! \binom{s}{k} \mathbb{E}\left[q_{x,n}^{k} | \mathcal{F}_{n-1}\right]  \mathcal{Z}\!\left( k, \delta, \theta \right) \Big]^{-1}.
\end{align}
Leveraging Lemma~1, we have
\begin{align}\label{equ: q_xn}
\mathbb{E}\left[q_{x,n}^{k} | \mathcal{F}_{n-1}\right] &= \mathbb{E}\left[ \chi_{\{ \mu^\Phi_{x, n} \leq \xi \}} + \left( \frac{\xi}{\mu^\Phi_{x, n}}\right)^k \chi_{\{ \mu^\Phi_{x, n} > \xi \}} | \mathcal{F}_{n-1}\right]
\nonumber\\
&= F^\Phi_{n-1}(\xi) + \int_{\xi}^1 \frac{\xi^k}{t^k} F_{n-1}^\Phi(dt)
\end{align}
where $\chi_{\{ \cdot\}}$ is the indicator function. By substituting \eqref{equ: q_xn} back into \eqref{equ: M_Yn}, and using the the Gil-Pelaez theorem for another time, we have the CDF of $\mu^\Phi_{x, n}$ given as
\begin{align}\label{equ: F_xn}
F^\Phi_{x, n}(u) &= \mathbb{P}\left( \mathbb{P}\left( \gamma_{x, n} > \theta | \Phi \right) < u \right)
= \mathbb{P}\left( Y^\Phi_{x,n} < \ln u \right)
\nonumber\\
&= \frac{1}{2} - \frac{1}{\pi} \int_{0}^{\infty} \frac{1}{\omega} \mathrm{Im}\left\{ u^{-j \omega} M_{Y^\Phi_{x, n}}(j \omega) \right\} d\omega.
\end{align}

Noet that $F^\Phi_{n}(u)$ appears on the left hand side of \eqref{equ: F_xn}, and $F^\Phi_{n-1}(u)$ appears on the right hand side. As $F^\Phi_{n}(u) \leq F^\Phi_{0}(u), \forall u \in [0, 1], n \in \mathbb{N}$, by the Dominant Convergence Theorem \cite{Bil:08}, we have $F^\Phi_{n}(u) \rightarrow F^\Phi(u)$, as $n \rightarrow \infty$, and the result follows.

\subsection{Proof of Theorem~\ref{thm: moment}} \label{appnd: Prf_moment}
We denote $\mathcal{F} = \cup_n \mathcal{F}_n$ to be the $\sigma$-algebra of all the queuing status at the steady state, and $q_x$ the corresponding active probability at a generic node $x$. As such, the $m$-moment of the conditional success probability can be computed as
\begin{align}
&\mathbb{E}\left[ \lim_{n \rightarrow \infty} \mathbb{P}\left( \gamma_{x, n} > \theta | \Phi \right)^m \right]
\nonumber\\
&=  \mathbb{E} \! \left[ \lim_{n \rightarrow \infty} \!\!\! \prod_{z \in \Phi \setminus x } \!\!\! \left( \frac{\mathbb{E}[q_{z}| \mathcal{F}_n ]}{1 \!+\! \theta \Vert x \Vert^{\alpha}/ \Vert z \Vert^{\alpha} } \!+\! 1 \!-\! \mathbb{E}[q_{z}| \mathcal{F}_n ] \right)^{m}  \right]
\nonumber\\
&= \Big[1 + \delta \sum_{k=1}^{m} \! \binom{m}{k} \mathbb{E}\left[q_{x}^{k} | \mathcal{F} \right]  \mathcal{Z}\!\left( k, \delta, \theta \right) \Big]^{-1},
\end{align}
and the result follows by using Lemma~1 and \eqref{equ: Meta_Grl} to the above equation.

\end{appendix}

\bibliographystyle{IEEEtran}
\bibliography{bib/StringDefinitions,bib/IEEEabrv,bib/howard_trff_schedule}

\end{document}